# McStas and Mantid integration

T R Nielsen[1], A J Markvardsen[2], and P Willendrup[1,3]

[1]European Spallation Source ERIC, Universitetsparken 1, 2100 Copenhagen Ø, Denmark

[2]ISIS Facility, Rutherford Appleton Laboratory, Chilton, Didcot, Oxfordshire, United Kingdom

[3]Physics Department, Technical University of Denmark, 2800 Kongens Lyngby, Denmark

## Abstract

McStas and Mantid are two well-established software frameworks within the neutron scattering community. McStas has been primarily used for simulating the neutron transport mechanisms in instruments, while Mantid has been primarily used for data reduction. We report here the status of our work done on the interoperability between the instrument simulation software McStas and the data reduction software Mantid. This provides a demonstration of how to successfully link together two software packages that otherwise have been developed independently, and in particular here show how this has been achieved for an instrument simulation software and a data reduction software. This paper will also provide examples of some of the expected future enhanced analysis that can be achieved from combining accurate instrument and sample simulations with software for correcting raw data. The main application of this work is to treat raw data collected at large scale neutron facilities.

## Introduction

The McStas [1,2] neutron ray-tracing simulation package is a versatile tool for producing accurate simulations of neutron scattering instruments at reactors and pulsed spallation sources. McStas is extensively used for design and optimization of instruments, virtual experiments, and user training. McStas is an abbreviation for Monte Carlo Simulation of triple axis spectrometers, but allows for describing and simulating any type of neutron scattering instrument.

The Mantid [3,4] project provides a framework that supports high-performance computing and visualisation of scientific data. Mantid has been created with the purpose to manipulate and analyse neutron and muon scattering data. Mantid is a short name for Manipulation and Analysis Toolkit for Instrument Data.

The purpose of the work presented here is to report on the work done on integrating McStas and Mantid, and the additional benefits which can be gained from using both these software packages for data treatment and analysis.

In many scientific domains excellent software exists for simulation and modelling. They are however often not applied directly against real raw data, but rather targeted at reduced data in the form of e.g. the differential scattering cross-sections $I(q,\omega)$ as a function of the transferred momentum Q and energy $\omega$. Additional reductions and transformations to correct raw data from instrument specific effect such as e.g. sample environment scattering, sample holder scattering contribution, and multiple scattering is often desirable in order to obtain a direct comparison between theory and experimental data. What a combination of McStas and Mantid can offer is an improved data procedure of getting "cleaner" $I(q,\omega)$ and potentially a better understanding of the raw data, by directly comparing such data with McStas simulations where the McStas sample may not attempt to describe the physics of a real sample fully, but just the parts which are known. This would give the scientist the opportunity to subtract the known physics from real data and to see what remains.

A number of large-scale neutron facilities use the NeXus data format [5] for storing neutron data. NeXus is derived from the Hierarchical Data Format (HDF) [6]. The details of how individual neutron facilities store neutron counts and metadata in NeXus files varies. However the NeXus application programming interface (API) is supported on multiple platforms and has library support for multiple programming languages [5]. McStas can write NeXus files and Mantid can read and write NeXus data files.

Just as the specific details of how individual neutron facilities store neutron counts and metadata in NeXus files varies so the McStas and Mantid NeXus files are also per se not compatible. However by a few adjustments, McStas NeXus data has, as part of this work, been formatted in a way that is compatible with Mantid and in line with how data are recorded at neutron spallation facilities.

At any neutron facility, detecting a neutron implies recording two pieces of information: the time of arrival, i.e. when a neutron is observed, and the detector pixel ID which uniquely identifies in which detector pixel the neutron is detected, i.e. where it was observed. This is the most detailed information that can be observed with current detector acquisition technology. Neutron data stored in this way are known as event data.

Hence McStas has been extended to output a NeXus event data format aligned with Mantid. To significantly speed up the neutron transport simulation McStas uses the concept of neutron rays and what is noted as 'weighted' neutron events. These can be naturally handled with Mantid, i.e. Mantid data reduction algorithms work with both weighted event data and 'normal' event data.

A totally realistic semi-classical simulation of neutron transport will require that each neutron is at any time either scattered or absorbed. In many instruments, only a very small fraction of the initial neutrons will ever be detected, and simulations of this kind will therefore waste much time in dealing with neutrons that never hit the relevant detector or monitor. An important way of speeding up calculations is to introduce a neutron "weight factor" for each simulated neutron ray and to adjust this weight according to the path of the ray. If e.g. the reflectivity of a certain neutron transport component is 10%, and only reflected neutron rays are considered later in the simulations, the neutron weight will be multiplied by 0.10 when passing this component, but every neutron is allowed to reflect in the component. In contrast, the totally realistic simulation of the component would require on average ten incoming neutrons for each reflected one. See the McStas manual for further details [1].

## Work carried out on McStas to enable exporting data to Mantid

From the beginning of the project, it was identified that the essential information that needs to be transferred from McStas to Mantid is a description of the instrument geometry in order for Mantid to perform a time-of-flight to wavelength analysis. This implies parsing information about a) where the neutron source is located, b) where the sample is located, c) where each individual detector pixel is located. Mantid has strong support for handling neutron event data, where each recorded neutron is given a time stamp and a pixel ID reflecting when and where the neutron was detected, and thus the outcome of a McStas simulation must provide such information. Finally McStas records histogram data, e.g. intensity as a function of wavelength, which need to be stored together with the event data and be read by Mantid for two reasons: 1) to ensure familiarity with the McStas user community. 2) To mimic e.g. how all monitor data at the Spallation Neutron Source in the USA and other time-of-flight facilities are recorded, i.e. as histogram data in order to save storage space. To implement such a solution, modifications were carried out in a number of key areas of McStas which we will describe below.

In McStas two types of simulation run-types are possible for different types of output. One can perform either a 'trace' or a 'simulation'. The so-called 'trace' uses the mcdisplay tool, which presents the instrument geometry, transferred to e.g. PGPLOT or Matlab graphics. As this type of run is aimed at presenting instrument geometry and looking for issues of connectivity of components and where neutrons are being lost, typically a low number of neutron rays are simulated. The so-called 'simulation', where a larger number of neutron rays are simulated through the instrument and monitor histograms of their parameters are gathered at specific points in the instrument geometry. Figure 1 shows the neutron ray-tracing view of a small angle neutron scattering (SANS) McStas simulation, while Fig. 2 shows the corresponding SANS scattering pattern recorded on the detector.

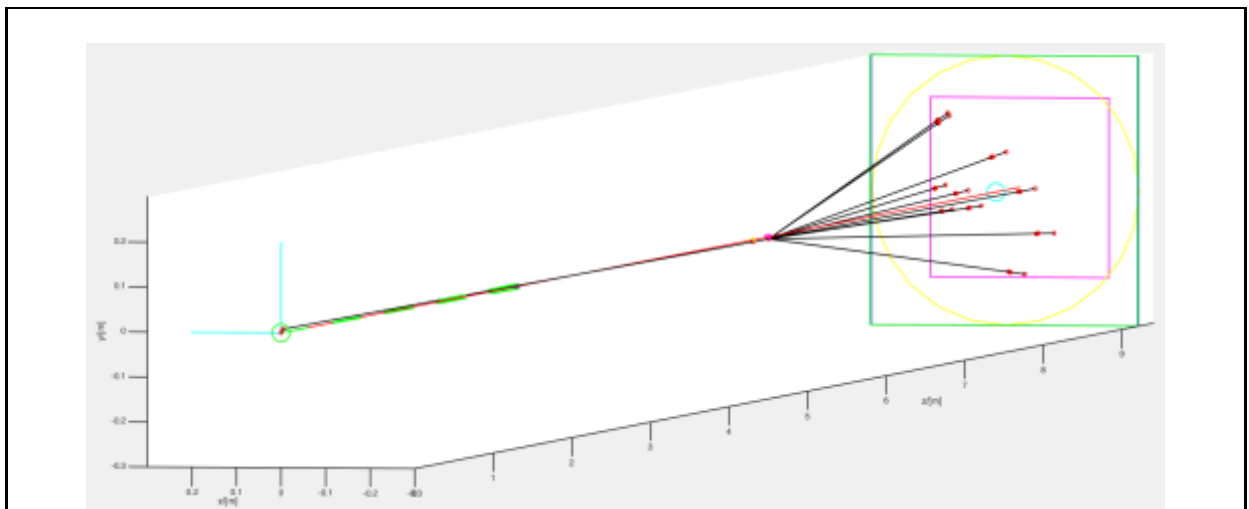

Figure 1: McStas trace mode. 3D visualisation of a virtual SANS experiment using the "templateSANS_Mantid" instrument. Neutron rays are scattered in a SANS pattern from hard spheres in dilute solution.

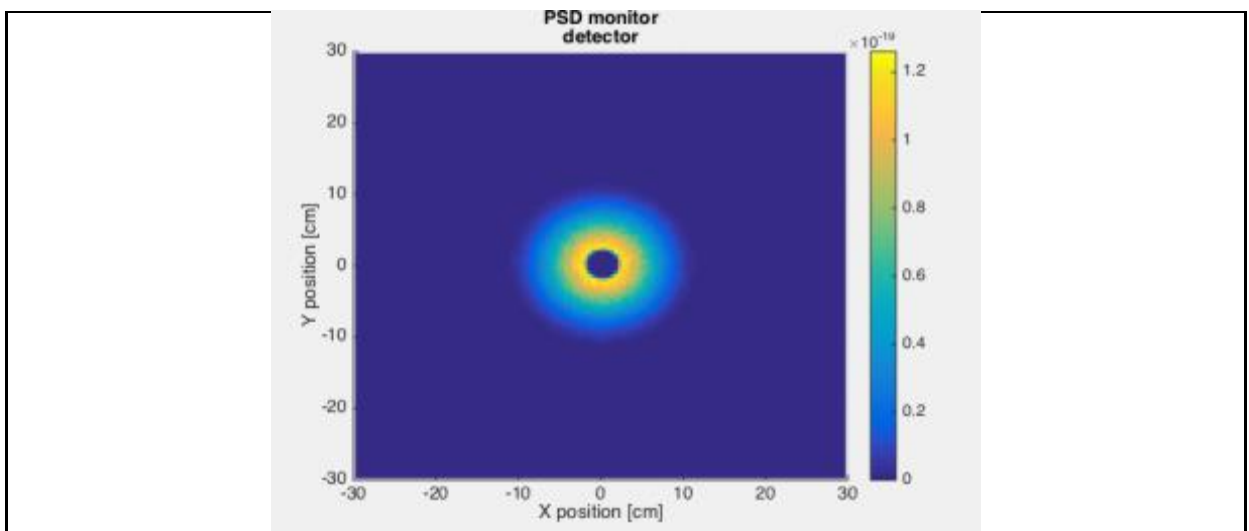

Figure 2: McStas simulation mode. Recorded SANS pattern histogram from a 2D position sensitive detector (PSD). The instrument model is "templateSANS_Mantid" and the sample is hard spheres in dilute solution.

As the McStas 'trace' mode is already capable of transferring geometrical information to various output formats, it was chosen as the basis for the generator that passes McStas geometry information to Mantid. The instrument description in Mantid is denoted Instrument Definition File (IDF) and is an xml file [7].

When run with the parameters 'mcdisplay --format=Mantid' an IDF is automatically generated from the components described in the McStas instrument file. If 'mcdisplay' is called with the '--complete' parameter, all geometry of all components is defined in the IDF for graphical representation in Mantid. In this way the IDF generated from McStas can be used to view the entire instrument in Mantid. By default the McStas-generated IDFs only list the source, sample and event detectors in accordance with the minimal information needed to lay out the instrument in Mantid. At present a detector cannot be tagged as a Mantid instrument monitor from a McStas generated IDF, hence currently histogram and event monitor data are treated separately as is demonstrated in the code example in Fig. 9. Hence, these key components are handled in a special way:

The location of the source is defined by naming a component entity 'sourceMantid'. That component does not need to correspond to the physical location of the source component in McStas. This option is needed for instruments with a curved guide, as Mantid currently calculates the distance between source and sample simply as the straight distance between those two points. So the source of an instrument with a curved guide should be placed with its orientation parallel to the *last* guide element, but translated backwards to define the correct length of the incoming flight path. Currently the McStas generated IDF does not handle the case where the flight path between the sample and the detector is not a straight line, including the handling of indirect geometry spectrometer instruments. This could be implemented by recording neutronic detector positions in addition to physical detector positions [7].

The location of the sample is defined by naming a component entity 'sampleMantid' and, as for the source component, this sample identification component does not need to correspond to the physical location of the sample component in McStas.

A number of Monitor_nD components are introduced in order to use the event detectors in Mantid. Each event monitor must be named 'nD_Mantid_k' where k is an integer counted from 0 and onwards if there are several detector banks.

A special option string that defines a rectangular panel detector, a cylindrical panel detector or an arbitrary-geometry detector where pixels are defined as surface elements from a Geomview OFF object file [8] is used to define the event detector for Mantid usage. The option strings for the three cases are listed below. The '*min*' identifier is used to define the numbering of pixels, i.e. the detector PIXEL_IDs for Mantid:

---

1) Rectangular detector:
options ="mantid square x limits=[-0.2 0.2] bins=128 y limits=[-0.2 0.2] bins=128, neutron pixel min=500 t, list all neutrons"

2) Cylindrical cut "banana" detector:
options="mantid banana, theta limits=[-28.0 135.38] bins=389, y limits=[-1.600 -0.600] bins=128, neutron pixel min=50000 t, list all neutrons"

---

Once an IDF has been generated using the mcdisplay utility, subsequent simulations with NeXus output will 1) embed the IDF file in the correct part of the NeXus tree for interpretation in Mantid, 2) store the incoming neutron events in event lists that can be interpreted by Mantid, 3) store ordinary McStas histogram data as histograms that can be interpreted by Mantid.

For completeness let us point out that in addition to the above-mentioned modifications the McStas NeXus output generation was slightly altered (code refactoring), as well as event output to lists being optimised, especially to enable NeXus and MPI parallelisation.

## Work carried on Mantid to enable importing McStas data

Mantid loads data into units called workspaces [9], and a loader, 'LoadMcStas' [10], was created for loading the McStas Nexus output files, a file format containing histogram (summary) data, event data as well as an IDF used for event data as detailed in the previous section. 'LoadMcStas' loads McStas histogram data into Mantid in workspaces of type Workspace2D which, once loaded, can be used in data treatments or simply be displayed and inspected using tools available in Mantid. The McStas event data are loaded into workspaces of type 'EventWorkspace' together with McStas' geometric description of the instrument; where during the loading phase each McStas detector event gets associated with the detector pixel ID detecting it. The result of this is that McStas event data are available in Mantid on an equal footing with 'EventWorkspace' workspaces created by loading event data measured on instruments at neutron scattering facilities such as ISIS and SNS. This means that all Mantid data treatment algorithms [11] can then be applied to McStas event data also.

As part of writing the loader, 'LoadMcStas' was optimised to work with the Nexus API to handle large datasets. Furthermore, a McStas instrument description may include complex geometry sources and sample descriptions. Mantid was extended to cope with such instrument descriptions in the same way as complex detector and monitor descriptions are handled.

Below Mantid instrument views are shown of a McStas-generated instrument description of IN5 at ILL, see Fig. 4, and McStas-generated instrument description for the planned LoKI instrument at the European Spallation Source ERIC, see Fig 5. The colours of the detector pixels show the integrated intensity of weighted McStas events.

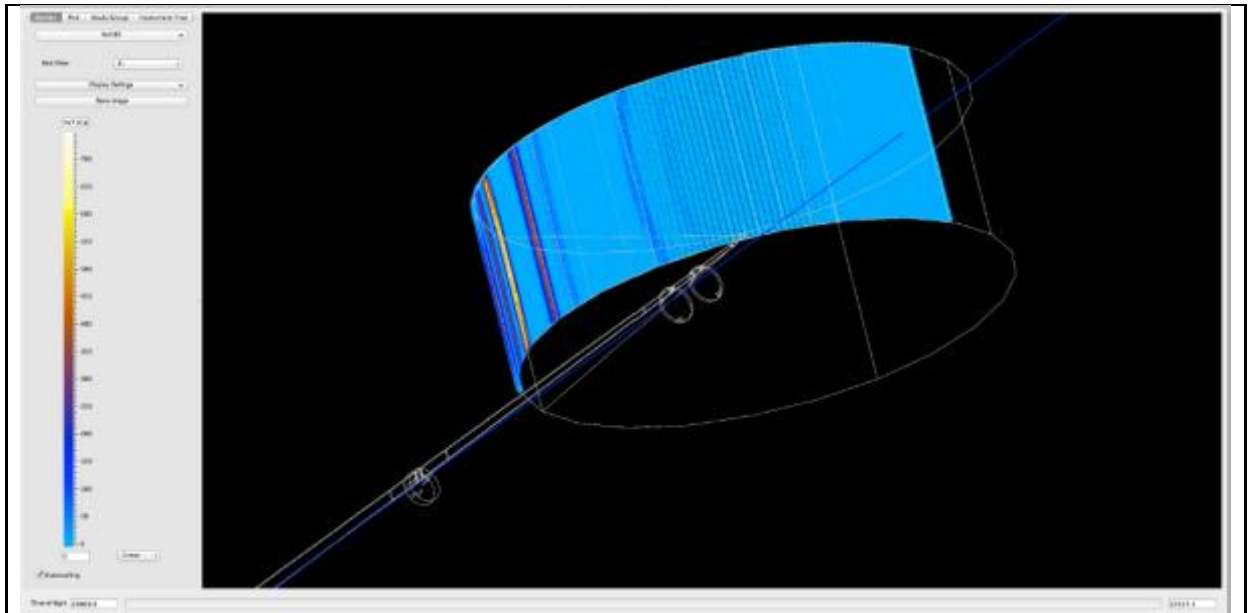

Figure 4: Mantid instrument view of a McStas-generated instrument description of IN5 at ILL. The detector description was generated using the cylindrical cut "banana" geometry. For the purpose of visualization Mantid shows all the event data as histogram intensity data displayed on top of the detector geometry.

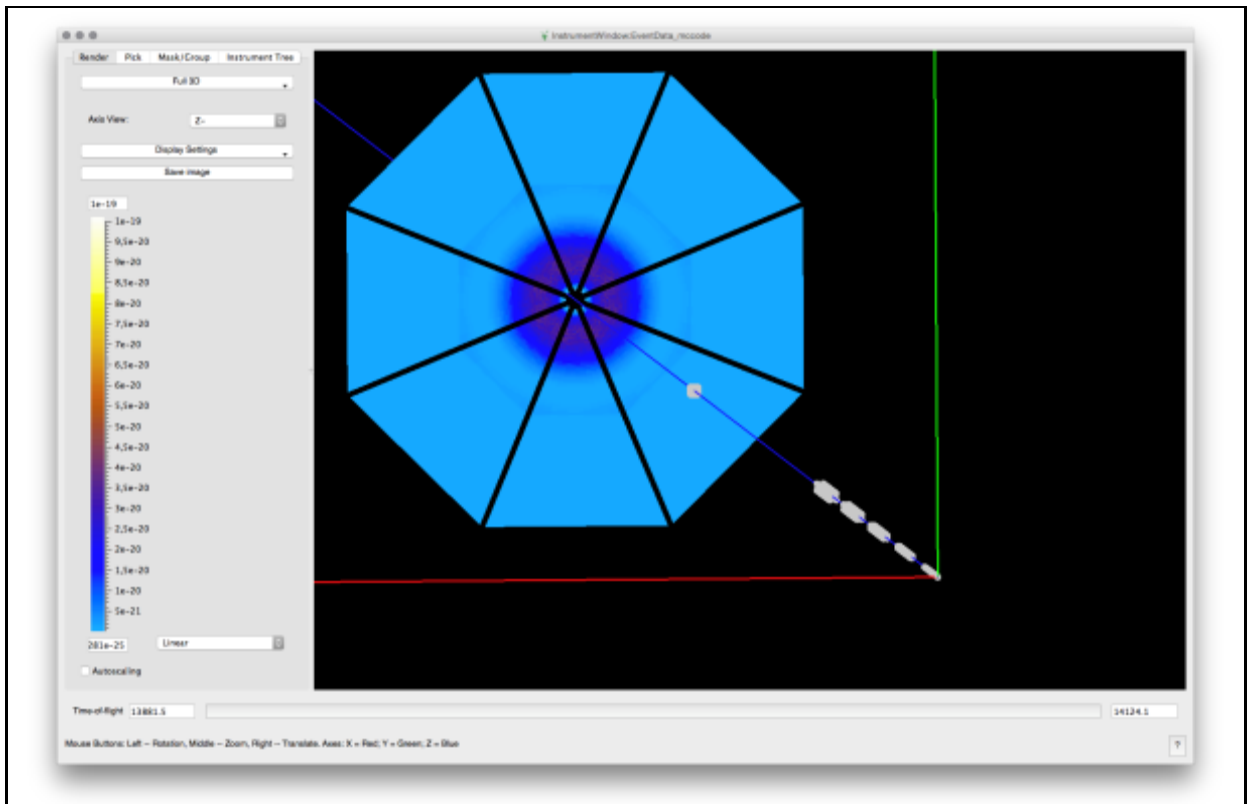

Figure 5: Mantid instrument view of a McStas-generated instrument description of LoKI at the European Spallation Source ERIC. Due to the new and unique LoKI detector design the McStas instrument description was generated using the Geomview OFF geometry, which is capable of describing arbitrary-geometry detectors.

A McStas event is stored in Mantid as a weighted event, which stores the McStas "weight factor" value for the simulated neutron. This may be compared to a Mantid non-weighted (raw) event that simply stores the pulse time and time-of-flight time of the event; where the pulse time is when a (neutron spallation) proton pulse occured in absolute time and time-of-flight is the time it took for the neutron to travel from the source to the detector. All McStas events are recorded with pulse time equal to zero.

From the point of view of data treatment Mantid handles weighted and raw events seamlessly. However, one thing to keep in mind is how errors are propagated [12] and treated in Mantid. For raw events there is not much to think about since a raw event simply stores the time of arrival of a neutron in a detector. When this information is gathered over time, signal error can be derived using normal counting statistics, i.e. assigning the square root of the measured counting statistics as an error estimate. McStas events record the "weight factor" and a Mantid weighted event stores an error estimate associated with this value. 'LoadMcStas' has been modified such that errors can either be set equal to one or assigned equal to the weight factor value (defaults to the latter) [13], which is consistent with the error estimation done in McStas.

## How to use McStas event data in Mantid

The McStas-Mantid data interface can read McStas event data into Mantid, from where on further data reduction and analysis can be performed. The purpose of this section is to demonstrate the methodologies, techniques and workflows used when combining McStas and Mantid. To demonstrate the usage of the algorithm, we set out to compare experimental data from the ISIS SANS2D instrument [14] to a McStas simulation of the same beam line, using the same beam and sample parameters. First we will describe the general syntax used for generating McStas event data for Mantid. Secondly we show a comparison between experimental data from ISIS SANS2D and a McStas simulation where both the experimental and simulated event data have been reduced to scattering intensity I(q) by Mantid reduction algorithms.

### How-to generate McStas event data for Mantid

McStas 2.x event data and the corresponding IDF for Mantid are generated as follows using these commands from a terminal. Here we use the McStas instrument file templateSANS_Mantid.instr which is distributed with the McStas framework:

```
Step 1. Compile McStas instrument code:
mcstas templateSANS_Mantid.instr --trace

Setp 2. Compile c code:
gcc -o templateSANS_Mantid.out  templateSANS_Mantid.c  -lm -DUSE_NEXUS -lNeXus

Step 3. Generate IDF:
mcdisplay templateSANS_Mantid.instr  --format=Mantid -n0

Step 4. Run simulation:
./templateSANS_Mantid.out  --format=Nexus
```
Figure 6: McStas code generation and compile methods.

## McStas event data conventions

For McStas to auto-generate an IDF the McStas instrument file must obey:

McStas instrument file name and the McStas defined name of the instrument must be the same:
E.g. templateSANS_Mantid.instr and "DEFINE INSTRUMENT templateSANS_Mantid(…. )"

In the McStas instrument file the source must be named "sourceMantid"
E.g. "COMPONENT sourceMantid = Source_simple(…. )"

In the McStas instrument file the sample must be named "sampleMantid"
E.g. "COMPONENT sampleMantid = Sans_spheres(…. )"

In the McStas instrument file the event monitors must be named "nD_Mantid_#"
E.g. "COMPONENT nD_Mantid_1 = Monitor_nD(…. )"

Figure 7: McStas event data conventions when used for Mantid

The McStas component monitor_nD must be called with the argument: options ="mantid square x limits=[-0.2 0.2] bins=128 y limits=[-0.2 0.2] bins=128, neutron pixel t, list all neutrons", where the number of bins and limits can be freely chosen.

## McStas simulations using the ISIS_SANS2d.instr file

A theory-experiment comparison is made for SANS investigation of spherical nanoparticles, with data measured at the ISIS SANS2D beam line. The sample can be characterized as follows: $SiO_2$ spherical nanoparticles that are ~ 150 Å in radius in $D_2O$, the spherical nanoparticles are characterised by a ~20% size variation, and the volume fraction of particles is just under 1% [15].

The correct McStas description of the neutron optics for ISIS SANS2D is entailed in the ISIS_SANS2d.instr file, which is included in the McStas neutron suite of instrument descriptions. The spherical nanoparticles are, as indicated above, characterized by a non-monodispersive size distribution. To account for such diversity in the scattering particles the standard McStas scattering components must be enhanced to describe a sample of spherical nanoparticles of different sizes. We enhance the McStas component "Sans_spheres.comp" such that it can describe an ensemble of spherical nanoparticles with a Gaussian size distribution with mean radius r and variance σ = PD × r where PD is the polydispersity fraction. The scattering from the modified "Sans_spheres.comp" component was verified by comparision with data from SasView [16]. Using the ISIS_SANS2d.instr and the modified "Sans_spheres.comp" files, McStas event data was generated as described in the sections above, and then imported into Mantid. The beam line optics and integrated neutron intensities at the detectors are shown in Fig. 8. The next step is to make a transformation of the McStas event data from real space coordinates (i.e., x, y, z, t) to scattering intensity as a function of momentum change, i.e. I(q).

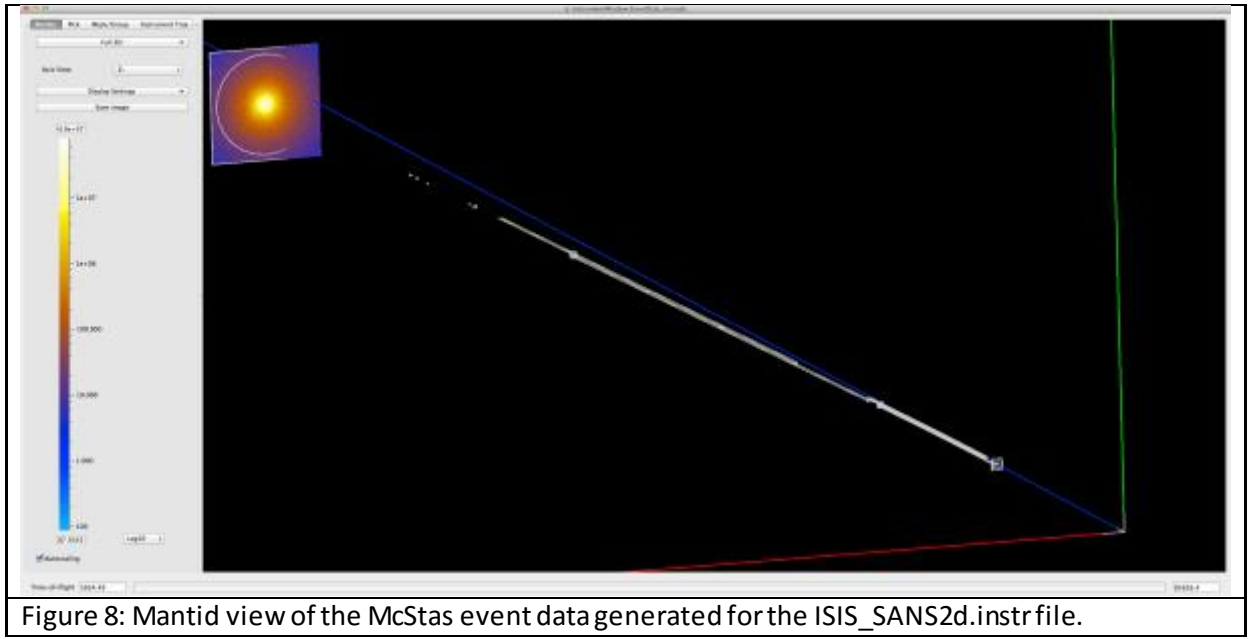

Figure 8: Mantid view of the McStas event data generated for the ISIS_SANS2d.instr file.

The workflow for processing SANS McStas event data in Mantid may then be as follows: transform from time-of-flight to wavelength, transform from wavelength to momentum change, and then finally normalize data with the incoming flux distribution. More explicitly this involves, in one example, the steps

1. Load McStas event data into Mantid, using the "Load" algorithm
2. Bin each event into a specified time grid, using the "Rebin" algorithm
3. Convert time of flight event data to wavelength, using the "ConvertUnits" algorithm
4. Bin each wavelength event to a specific wavelength grid, using the "Rebin" algorithm
5. Clone McStas pre sample wavelength monitor, using the algorithm "CloneWorkspace"
6. Convert to histogram, using the algorithm "ConvertToHistogram"
7. Rebin to same wavelength grid, using the "Rebin" algorithm
8. Create monitor for latter normalization, using the algorithm "CreateWorkspace"
9. Reduce data to I(q), using the algorithm "Q1D"

The workflow can be put in to a small python script which is executed from within Mantid in order to evaluate I(q) for McStas event data.

```
# load both event and summary data into a the Python tuple ws.
# The loaded workspaces can then either be accessed by there name using mtd[''] notation
# or the order they are loaded into Mantid, where ws[0] is the first loaded event dataset
ws       = Load('mccode.h5')

# Rebin the event data into a grid of equal size 10 and convert to wavelenght
sample = Rebin(ws[0],'10', False)
sample = ConvertUnits(sample, 'Wavelength')

# Rebin wavelength converted data onto grid from 0.5 to 15.0 in steps of 0.005
binning  = '0.5,0.005,15.0'
sample = Rebin(sample, binning)
```

```
# Clone monitor data, which are loaded into Mantid as so-called point data
# and needs to be converted to histogram data and binned to the same binning
# as used for the sample data
mcstasMonitor  =  CloneWorkspace(mtd['lmonitor3.dat_ws'])
mcstasMonitor  =  ConvertToHistogram(mcstasMonitor)
mcstasMonitor  =  rebin(mcstasMonitor, binning)
# Translate McStas wavelength label into Mantid wavelength unit (the need for this
# workaround scheduled to be removed for Mantid v 3.7)
monitor              = CreateWorkspace(mcstasMonitor.dataX(0),
                            mcstasMonitor.dataY(0),
                            mcstasMonitor.dataE(0),
                            UnitX='Wavelength')

# calculate I(q), normalising the sample data with the monitor spectrum and with
# a q-binning from 0.001 to 0.1 in steps of 0.001 and then with steps of 0.01 till 0.7
reduced_norm = Q1D(sample, '0.001,0.001,0.1,0.01,0.7',WavelengthAdj='monitor')
```

Figure 9: Python script for Mantid reduction of McStas event data.

Using the Mantid Python script shown above in Fig. 9 the McStas event data are reduced and the corresponding scattering intensity I(q) is shown in Fig. 10. For comparison the experimental data are also shown. The experimental data are reduced to I(q) using the SANS interface in Mantid.

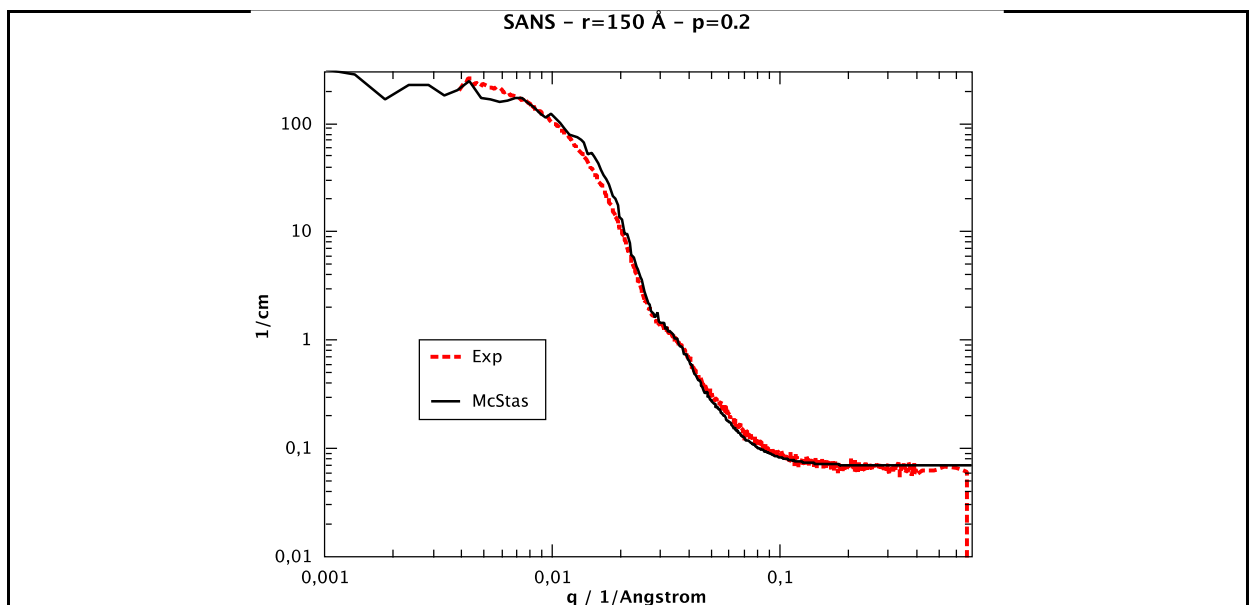

Figure 10: Comparison of rescaled scatting intensity I(q) derived from the experimental data and a McStas simulation.

There seem to be good agreement between the simulated McStas data and the measured data, which validates the underlying workflow for the Mantid reduction of McStas event data.

## Example of expected benefits of joint McStas/Mantid analysis

In this section we discuss some of the expected benefits of a joint usage of McStas and Mantid for analysing neutron scattering experiments.

## Using existing algorithms already developed for Mantid

Once McStas event data have been loaded into Mantid, all of Mantid analysis methods [10] can be applied to the dataset. This makes already developed and tested methods from Mantid directly applicable to McStas data, and will in turn make the analysis faster as the users do not have to reinvent and re-implement reduction and analysis methods. As an example of this we consider a McStas simulation of a SANS experiment. Compared to the previous SANS example we use here a McStas scattering kernel, which gives an asymmetric scattering pattern. We considered nano cylinders in a solution. For the purpose of demonstration, all nano cylinders are orientated in the same direction with respect to the beam axis. The analysis runs very similarly to the previous section, only now using the 2D reduction 'Qxy' method instead of 'Q1D'. From a McStas viewpoint, this is a clear improvement, where the users have to focus the neutron transport description, and let Mantid take care of the reduction. Figure 11 shows the integrated intensity recorded on a square detector, while Fig. 12 shows the 2D scattering intensity as a function of qx and qy. The reduced scattering intensity I(qx,qy) based on the McStas-Mantid workflow coincides with the analytic expression for scattering cross-section [16].

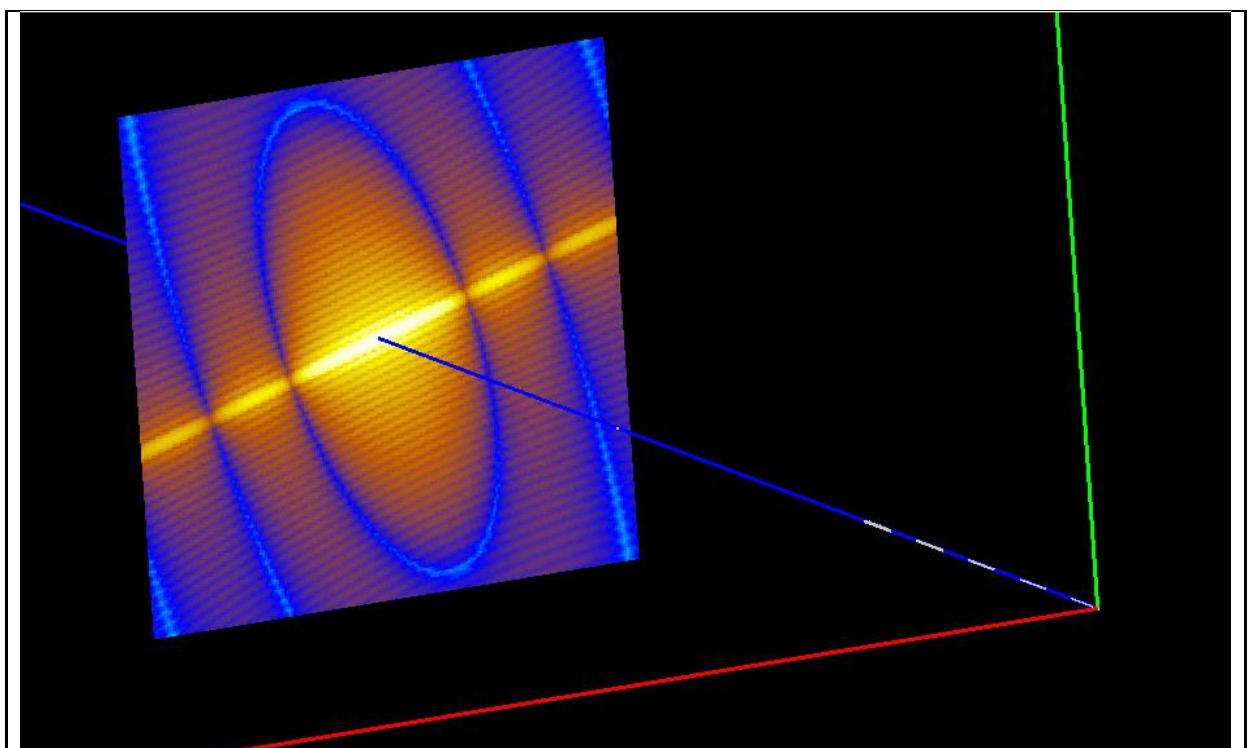

Figure 11: Mantid view of the McStas event data generated for a virtual SANS experiment using the "templateSANS_Mantid" instrument. The scattering sample consists of hard cylinders in a dilute solution. All nano particles have the same orientation.

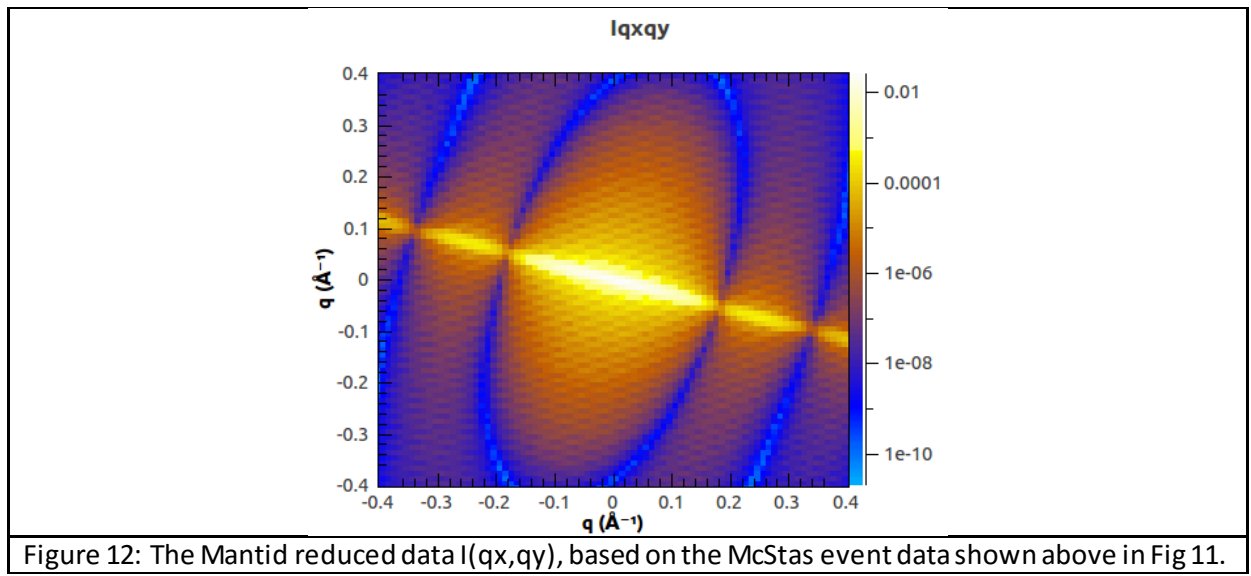

Figure 12: The Mantid reduced data I(qx,qy), based on the McStas event data shown above in Fig 11.

## Pre and post analysis

Current best practice in neutron scattering experiments is to reduce and analyse data once the experiment is done. The experimental scientists would then have to do a post analysis of the reduced data. The outcome of such an data analysis will be reflected by the experimental settings and how the experiment was carried out. If the data analysis is inconclusive and the experimental parameter settings turn out to have been ill defined, new experimental time is needed for further studies. By using McStas and Mantid such situations can be reduced, and beam time can be better utilized by setting up virtual experiments prior to doing actually experiments. Mantid has already some tools available for estimating parameters for beam time; see for instance Ref. [17], but none of them are based on actually neutron transport calculations. Thus, the easy interoperability between McStas and Mantid facilitates running virtual experiments where users prior to an experiment can plan the experimental run in detail; e.g. check the experiment parameter settings, as well as being used to preform training exercises for a neutron scattering instrument.

McStas and Mantid can also be used for improving data post processing methods. One could envisage a workflow as follows

- reduce experimental data with Mantid
- use Mantid and/or 3rd party analysis software tool to derive physical parameters
- start a new McStas simulation based on the above derive physical parameters
- reduce the new McStas event data with Mantid
- use Mantid and/or 3rd party analysis software tool to derive physical parameters
- repeat steps above ..........

Since each neutron ray can be tagged in McStas according to where and how it was scattered, different contributions to the total scattering pattern may be inferred or visualized. For example: multiple scattering from the sample environment and sample, multiple scattering from within the sample, or discrimination between coherent and incoherent scattering can be identified, see Ref. [18] for further details.

## Calculate multiple scattering

Combining McStas and Mantid has the potential of providing a generic way of calculating the multiple scattering signal from a sample. This of course requires that McStas has good models for multiple scattering within a sample and/or any interaction with sample environment and beam windows. Assuming the multiple scattering signal is not the dominating scattering signal from the sample then the following joint use of McStas and Mantid could be used to estimate the multiple scattering signal:

Step 1. Using Mantid calculate an $I`(q, \omega)$ from a raw dataset collected on beamline instrument XX. In this treatment process no attempt should be made to correct for multiple scattering.

Step 2. Run a McStas simulation setup for instrument XX and with a sample description that mimics the real sample as far as possible including setting $I(q, \omega)$ for the sample equal to $I`(q, \omega)$. Further set this simulation up so neutrons which have been scattered more than once in the sample get stored separate from neutron which have been scattered once; call these two simulated results SIM_MulScat and SIM_SingleScat respectively.

Step 3. As a test of the McStas simulation, from the SIM_SingleScat dataset, $I`(q, \omega)$ must be recovered after data treatment. Then using the exact same data treatment on SIM_MulScat a 'first order' approximation of the multiple scattering signal for this sample is obtained. If we call this $\Delta(q,\omega)$, then correct $I`$ for multiple scattering using $I`` = I` - \Delta$.

The purpose of steps 1 to 3 is not to calculate with 100% absolute accuracy the multiple scattering signal, but to give the user something to work with for cases where multiple scattering is known to be present but the user does not know how to otherwise estimate this contribution. Note that this will not recover information from sharp features that have been smeared away by multiple scattering. For efficiency of repeated calculations, the McStas instrument may be modelled from just before the sample, using a previously stored incident spectrum.

## Detector calibration

One suggestion [19] is that for instruments that contain banks of detectors, it may be more efficient to keep the detector positions description in McStas according to the engineering positions, meaning, for example, on a regular grid of equally spaced grid positions, rather than individual calibrated positions. If the calibration positions only differ little from the engineering positions then it may be expected that the McStas simulation output will only differ little depending on whether a calibrated or non-calibrated instrument was used, and that for the purpose of subsequent analysis can be ignored. The added benefit of using engineering positions with McStas is that it simplifies the instrument description and reduce the execution time of the simulation. A file containing detector calibration information (for example in the form of a Mantid parameter file, see [7]) could be applied after McStas data have been loaded into Mantid (or as a possible future alternative, McStas could be enhanced to allow such a calibration file as part of its instrument description, save it in the McStas Nexus generated output file, which then could be automatically read by Mantid during the loading of such a Nexus file).

## Take instrument design one step further

Run a McStas simulation followed by a Mantid reduction of the simulated data as if these were real data. An instrument design can then be tested to see if it can deliver its designed science programme to a more detailed level. In a simple case, a McStas simulation followed by a Mantid reduction may be run with or without a sample. The latter could e.g. be used as a basis for calculating the instrument resolution and the former to give more information as to what science can be retrieved from data collected on samples on an instrument.

## An option for further integration

As part of this work, McStas has been enhanced to automatically create Mantid IDFs from McStas instrument files. It would be advantageous if Mantid could be enhanced to generate McStas instrument files directly from all the IDF descriptions included in the Mantid framework. For example the Mantid instrument description language [7] could be extended to understand a syntax like:

```
<component type="DiskChopper" name="Chopper1"
  <location x="10.6" y="0.0" z="0.0" />
  <external-program-specific name="McStas" version="2.0">

    Some code specific to a version of McStas

<external-program-specific/>

</component>
```

Figure 13: Pseudo code for common McStas Mantid IDF

The black text above describes an instrument component of type 'DiskChopper' located at (10.6, 0, 0) and the red text is new syntax associating some external program-specific code, in this case for a version 2.0 of McStas with the component named 'Chopper1'. With information of this type a parser for converting a Mantid IDF to a McStas instrument file can be written. The use of <external-program-specific> code should be kept to a minimum. McStas instrument description is very rich and in some cases it likely makes no sense to try and enhance the Mantid instrument description language to accommodate all McStas specific language features. Different versions of McStas may have different specialised features it supports, hence the idea of the 'version' attribute, if specified, tells for which version of the McStas (in this case) specified code is known to work.

<external-program-specific> information may also be allowed to be specified in Mantid instrument parameter files [7]. This may be done to separate out such information into a separate file. For example the information in the red text above could alternatively be stored in an instrument parameter file as:

```
<component-link name=" Chopper1">

  <external-program-specific name="McStas" version="2.0">

    Some code specific to a version of McStas

  <external-program-specific/>

</component-link>
```

Figure 14: Pseudo code for common McStas Mantid IDF

Scenarios where translating a Mantid IDF to a McStas IDF could be useful include: the Mantid IDF contains the full instrument guide information, perhaps inherited initially from a McStas IDF description. Users would then have the option to modify the Mantid IDF and run McStas simulations on such modified instrument descriptions.

Detailed information about the incident beam optics and choppers etc. is as of this writing rarely included in Mantid IDFs, for the simple reason that such information so far has not been required for data reduction. However, if real data from an instrument monitor situated before the sample is available, the monitor position and data could be used to create a McStas source. For this scenario there may be no need for <external-program-specific> information. Note a McStas simulation that does not need to ray-trace neutrons through a guide can be expected to run considerably faster than a simulation from a fully described instrument.

## Summary

We have described our work done on the interoperability between the instrument simulation software McStas and the data reduction software Mantid. Specifically we have shown how to link McStas generated event data to Mantid's reduction algorithms. This facilitates new options for analysis tools to the users. Once McStas event data have been loaded into Mantid, all of Mantid analysis methods can be applied to the data set. This makes already developed and tested methods from Mantid directly applicable to McStas data, and will in turn make the analysis faster as the users do not have to reinvent and re-implement reduction and analysis methods. Furthermore, we have discussed possible benefits users could be expected to obtain by using both McStas and Mantid for data analysis of neutron scattering experiments directly based on the methods described in this work.

## Acknowledgements

We would like to acknowledge Garrett E. Granroth (SNS, Oak Ridge National Laboratory, USA), Martyn Gigg and Nick Draper (Tessella Ltd., Abingdon, Oxfordshire, UK). Emanual Farhi (ILL, Grenoble, France) for work/insight on Monitor_mD. Richard Heenan and Sarah Rogers (ISIS facility, Oxfordshire, UK and European Spallation Source ERIC) for helping with providing a suitable first test dataset and very useful discussions. Andrew Jackson and Kalliopi Kanaki (European Spallation Source ERIC) for discussion related to detector designs, and Jon Taylor (European Spallation Source ERIC) for discussion on multiple scattering.

## References

[1] http://www.mcstas.org; Willendrup P. Farhi E. Lefmann K. McStas 1.7 - A New Version of the Flexible Monte Carlo Neutron Scattering Package, Physica B: Condensed Matter. Volume 350, Issues 1–3, Supplement, 15 July 2004, Pages E735–E737 / doi:10.1016/j.physb.2004.03.193

[2] Willendrup, P.; Farhi E.; Knudsen E.; Filges U.; Lefmann K; McStas: past, present and future. Journal of Neutron Research 17, 2014 pp. 35-43 / http://content.iospress.com/articles/journal-of-neutron-research/jnr004

[3] http://www.mantidproject.org; O. Arnold, et al., Mantid—Data analysis and visualization package for neutron scattering and μSR experiments, Nuclear Instruments and Methods in Physics Research Section A, Volume 764, 11 November 2014, Pages 156-166 / http://dx.doi.org/10.1016/j.nima.2014.07.029

[4] Mantid (2013): Manipulation and Analysis Toolkit for Instrument Data.; Mantid Project.http://dx.doi.org/10.5286/SOFTWARE/MANTID

[5] http://www.nexusformat.org

[6] https://www.hdfgroup.org/HDF5/

[7] http://www.mantidproject.org/IDF

[8] http://www.geomview.org/docs/html/OFF.html

[9] http://www.mantidproject.org/Workspace

[10] http://www.mantidproject.org/LoadMcStasg

[11] http://www.mantidproject.org/Algorithm

[12] http:www.mantidproject.org/ErrorPropagation

[13] Private communication Garrett E. Granroth

[14] "Small Angle Neutron Scattering using SANS2d", R.K.Heenan, S.E.Rogers, D.Turner, A.E.Terry, J.Treadgold & S.M.King, Neutron News 22(2011)19-21

[15] Private communication Richard Heenan and Sarah Rogers

[16] http://www.sasview.org

[17] http://docs.mantidproject.org/nightly/interfaces/DGSPlanner.html

[18] Private communication E. Farhi

[19] Private communication Garrett E. Granroth